\documentclass[twocolumn,amsmath,amssymb,aps]{revtex4}
\usepackage{graphicx}
\usepackage{bm}
\usepackage{amssymb} 
\usepackage{dcolumn}
\usepackage{wasysym}

\begin{document}

\title{Cooperative emission of light by an ensemble of dipoles near a metal nanostucture: The plasmonic Dicke effect}

\author{Vitaliy N. Pustovit$^{a,b}$ and Tigran V. Shahbazyan$^{a}$}

\affiliation{$^{a}$Department of Physics, Jackson State University, Jackson, MS
  39217, USA}

\affiliation{$^{b}$Laboratory of Surface Physics, Institute of Surface
  Chemistry, Kyiv 03164, Ukraine} 

\begin{abstract}
We identify a new mechanism for cooperative emission of light by an ensemble of $N$ dipoles near a metal nanostructure supporting a surface plasmon.The cross-talk between emitters due to virtual plasmon exchange leads to a formation of three plasmonic super-radiant modes whose radiative decay rates scales with $N$, while the total radiated energy is \emph{thrice} that of a single emitter. Our numerical simulations indicate that the plasmonic Dicke effect survives non-radiative losses in the metal.
\end{abstract}


\maketitle


Radiation of a dipole near a metal nanostructure supporting surface plasmon (SP) is attracting renewed interest due to possible  biosensing applications \cite{lakowicz-ab01}. While early studies mainly focused on fluorescence of molecules near rough metal films \cite{brus-jcp80}, recent advances in near-field optics and in chemical control of molecule-nanostructure complexes spurred a number of experiments on single metal nanoparticles (NP) linked to dye molecules \cite{feldmann-prl02,lakowicz-jf02,feldmann-nl05,novotny-prl06,sandoghdar-prl06,sandoghdar-nl07} or semiconductor quantum dots \cite{gueroui-prl04}. Emission of a photon by a dipole-NP complex involves two competing processes: enhancement due to resonance energy transfer (RET) from an excited dipole to a SP \cite{moskovits-rmp85}, and quenching due to decay into optically-inactive excitations in the metal \cite{silbey-acp78}. These decay channels are characterized by radiative, $\Gamma^{r}$, and non-radiative, $\Gamma^{nr}$, decay rates, respectively, and their balance is determined by the separation, $d$, of the emitter from the metal surface \cite{nitzan-jcp81,ruppin-jcp82}. The emission is most enhanced at some optimal distance, and is quenched close to the NP surface due to the suppression of quantum efficiency, $Q=\Gamma^{r}/\left (\Gamma^{r}+\Gamma^{nr}\right )$, by prevalent non-radiative processes. Both enhancement and quenching were widely observed in fluorescence experiments on Au and Ag nanoparticles \cite{feldmann-prl02,lakowicz-jf02,feldmann-nl05,novotny-prl06,sandoghdar-prl06,sandoghdar-nl07}. In recent \emph{single-molecule} measurements \cite{novotny-prl06,sandoghdar-prl06,sandoghdar-nl07}, the distance dependence was in a good agreement with \emph{single-dipole-NP} models \cite{nitzan-jcp81,ruppin-jcp82}, prompting proposals for a NP-based nanoscopic ruler \cite{sandoghdar-nl07}.

\begin{figure}[tb]
  \centering
  \includegraphics[width=0.8\columnwidth]{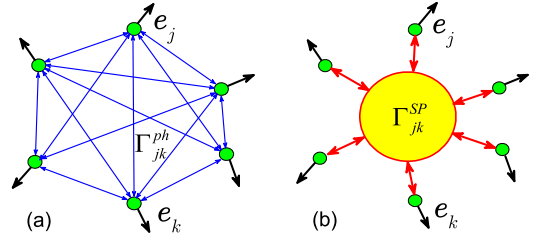}
  \caption{\label{fig:dicke} (Color online) Radiative coupling of emitters in free space (a) and plasmonic coupling of emitters near a NP (b).}
  \end{figure}

In this Letter, we identify a novel mechanism in the emission of light by an \emph{ensemble} of dipoles located near a nanostructure supporting a localized SP. A typical setup would involve, e.g., dye molecules \cite{feldmann-prl02,lakowicz-jf02,feldmann-nl05} or quantum dots \cite{gueroui-prl04} attached to a metal NP via DNA linkers. Namely, we demonstrate that RET between individual dipoles and SP leads to a \emph{cross-talk} between the emitters. As a result, the emission of a photon becomes a \emph{cooperative} process involving all dipoles in the ensemble. This \emph{plasmonic} mechanism of cooperative emission is analogous to the Dicke effect for $N$ radiating dipoles in free space, confined within a volume with characteristic size smaller than the radiation wavelength $\lambda$ \cite{dicke-pr54,andreev-book,scheibner-naturephys07}. In that case, the cooperative emission is due to photon exchange between the emitters that gives rise to super-radiant (SR) states with total angular momentun 1 and enhanced radiative decay rate  $\sim N\Gamma_{0}^{r}$, where $\Gamma_{0}^{r}$ is the decay rate of an \emph{isolated} dipole. In contrast, in plasmonic systems, the dominant coupling mechanism between dipoles is \emph{SP exchange}, i.e., excitation of a virtual SP in a nanostructure by an excited dipole followed by its absorption by another dipole, rather than direct radiative coupling [see Fig.~\ref{fig:dicke}]. Such a SP-induced coupling leads to the formation of \emph{plasmonic SR} states that dominate the emission of a photon.  Importantly, because the nanostructure acts as a \emph{hub} that couples nearby and remote dipoles with about equal strengths, the SP-induced cross-talk is more uniform throughout the ensemble, as compared to the radiation coupling, resulting in a more efficient hybridization and hence cooperative emission.

The usual photonic Dicke effect can be suppressed by internal non-radiative processes in molecules or by their energy exchange with the environment. For example, when an ensemble of emitters is located nearby a metal nanostructure, the photon exchange processes are largely quenched by ohmic losses in the metal. In contrast, as we demonstrate in this paper, the SP exchange mechanism is not significantly affected by quenching up to very small distances, and thus provides the main channel for cooperative emission in plasmonic systems.  Specifically, we show that for an ensemble of $N$ dipoles distributed in a solid angle around a metal NP, there are \emph{three} plasmonic SR states with radiative decay rates $\gamma_{\mu}^{r}\simeq N\Gamma^{r}/3$. Furthermore, in a wide range of dipole-NP distances, their \emph{non-radiative} decay rates also scale as $\gamma_{\mu}^{nr}\simeq N\Gamma^{nr}/3$, so that the SR \emph{quantum efficiencies} are essentially the \emph{same} as those of individual dipoles near a NP. As a result, the total energy radiated by an ensemble, $W$,  is only \emph{thrice} that radiated by a single dipole near a NP, $W_{0}$:
\begin{equation}
\label{energy}
W\simeq 3(\hbar kc/4)Q=3W_{0},
\end{equation}
where $k$ and $c$ are wavevector and speed of light, and the remaining energy is dissipated in a NP via sub-radiant states.  The fact that, in plasmonic systems, radiated energy of an ensemble is nearly independent on its size could allow unambiguous determination of single-molecule decay rates in situations when a large but uncertain number of molecules participate in the emission. 

\textit{Theory}---We consider a system of $N$ emitters, e.g., fluorescing molecules, with dipole moments $ {\bf d}_{j} = d_{j} {\bf e}_{j}$, where $d_{j}$ and  ${\bf e}_{j}$ are their magnitudes and orientations, respectively, located at  positions ${\bf r}_{j}$ around a spherical NP of radius $R$ in a dielectric medium with its center at origin.  We assume incoherent emission, i.e., molecules initially excited by a laser pulse, subsequently relax through internal transitions before emitting a photon, and adopt classical model of Lorentz oscillators with random initial phases. The frequency-dependent electric field, ${\bf E}({\bf r},\omega )$, created by all dipoles in the presence of a NP, satisfies Maxwell's equation
\begin{equation}
\label{maxwell}
\frac{\epsilon ({\bf r},\omega )\omega^{2}}{c^2}  {\bf E}({\bf r},\omega )-{\bm \nabla} \times {\bm \nabla} \times {\bf E}({\bf r},\omega )  = -
\frac{4\pi i \omega}{c^2} {\bf j}({\bf r},\omega ), 
\end{equation}
where dielectric permittivity $\epsilon({\bf r},\omega )$ is that of the metal inside NP, $\epsilon(\omega)$, for $r<R$, and that of the outside dielectric, $\epsilon_{0}$, for $r>R$. Here ${\bf j}({\bf r},\omega )= -i\int_{0}^{\infty}e^{i\omega t} {\bf j}(t) dt$, is the Laplace transform of dipole current ${\bf j}(t)=q\sum_{j} \dot{d}_{j}(t) {\bf e}_{j} \delta ({\bf r}-{\bf r}_{j})$, where dipole displacements are driven by the electric field at dipoles positions,
\begin{equation}  
\label{displ}      
\ddot {d}_{j} + \omega_{0}^2 {d}_{j} = \frac{q}{m} {\bf E} ({\bf r}_{j},t) \cdot {\bf e}_{j},
\end{equation}
with the initial conditions ${\bf d}_{j}=d_{0} {\bf e}_{j} \sin \varphi_{j}$, $\dot {\bf d}_{j}=\omega_{0}d_{0} {\bf e}_{j} \cos \varphi_{j}$, and ${\bf E}=0$ for $t=0$ (dot stands for time-derivative). Here $\omega_{0}$, $q$, $m$, and $\varphi_{j}$ are oscillators frequency, charge, mass, and initial phase, respectively ($\omega_{0}=\hbar q^{2}/md_{0}^{2}$). Closed equations for $d_{j}(\omega)$ are obtained by Laplace transforming Eq.~(\ref{displ}) with the above initial conditions and then eliminating ${\bf E}$ from Eqs.~(\ref{maxwell},\ref{displ}) \cite{shahbazyan-prb00}. The latter can be expressed via normalized displacements,  $v_{j}(\omega)=d_{j}(\omega)/d_{0} - i\left (\omega_{0}/\omega^2\right ) \cos \varphi_{j} - \omega^{-1}\sin \varphi_{j}$, and $v_{0j} = -i\left (\omega_{0}^3/\omega^2\right ) \cos \varphi_{j} - \left (\omega_{0}^2/\omega\right ) \sin \varphi_{j} $, as 
\begin{equation}
\label{electric}
{\bf E}({\bf r},\omega ) = - \frac{4\pi d_{0}q \omega_0^2}{c^2} \sum_{j}{\bf G}({\bf r},{\bf r}_{j},\omega) \cdot {\bf e}_{j} v_{j}, 
\end{equation}
where ${\bf G}({\bf r},{\bf r}',\omega)$ is the electric field Green diadic in the presence of NP. For the photon frequency close to those of dipoles, $\omega \approx \omega_{0}$, we arrive at the following system,
\begin{equation}
\label{system}
\sum_{k} \Bigl[(\omega_{0}-\omega ) \delta_{jk} +  \Sigma _{jk}\Bigr] v_{k} = 
\frac{v_{0j}}{2\omega_{0}}=\frac{-i}{2} e^{-i\varphi_{j}},
\end{equation}
where the complex \emph{self-energy matrix}, $\Sigma _{jk}$, is given by
\begin{equation}
\label{self}
\Sigma _{jk}(\omega)= - \frac{2\pi q^2 \omega_{0}}{m c^2} {\bf e}_{j}\cdot  {\bf G}(r_{j},r_{k};\omega) \cdot {\bf e}_{k}.
\end{equation}
The system (\ref{system}) determines \emph{eigenstates} of $N$ emitters coupled to each other via radiation field \emph{and} electronic excitations in a NP. For $|{\bf r}_{j}-{\bf r}_{j}|\ll \lambda$, we can use the near-field expansion of the Mie theory Green diadic ${\bf G}(r_{j},r_{k};\omega)$ \cite{ruppin-jcp82} for calculation of $\Sigma _{jk}$. The self-energy is dominated by imaginary part that contains the SP resonance. The details will be given elsewhere \cite{to-be}, and the \emph{decay matrix}, $\Gamma_{jk}=-{\rm Im} \Sigma_{jk}$, is a sum of radiative and nonradiative terms, $\Gamma_{jk}=\Gamma_{jk}^{r}+\Gamma_{jk}^{nr}+\delta_{jk}\Gamma_{0}^{nr}$, where 
\begin{align}
\label{width}
\Gamma_{jk}^{r}
&= \Gamma_{0}^{r} \left[({\bf e}_{j} \cdot {\bf e}_{k})-\alpha'_{1}
\left (K_{jk}^{(1)} + h.c.\right ) + |\alpha_{1}|^2 T_{jk}^{(1)}\right],
\nonumber\\
\Gamma_{jk}^{nr}
&=\frac{3\Gamma_{0}^{r}}{2k^{3}} \sum_l \alpha''_l T_{jk}^{(l)},
\end{align}
and $\Gamma_{0}^{nr}$ accounts for internal molecular transitions ($\delta_{jk}$ is Kronecker symbol). Here $\Gamma_{0}^{r}=2d_{0}^{2} k^{3}/3\hbar \epsilon_{0}$ is dipole radiative decay rate ($k=\sqrt{\epsilon_{0}}\omega/c$ is wavevector), $\alpha_{l}(\omega)=\alpha'_{l}(\omega)+i\alpha''_{l}(\omega)=\frac{R^{2l+1}\left [\epsilon(\omega) -\epsilon_0\right ]}{\epsilon(\omega) +(1+1/l)\epsilon_0}$ are $l$-pole nanoparticle polarizabilities,  and matrices $T_{jk}^{(l)}$ and $K_{jk}^{(l)}$ are
\begin{align}
\label{TK}
T_{jk}^{(l)}
&=\frac{4\pi}{2l+1} \sum_{m=-l}^{l} [{\bf e}_{j} \cdot {\bm \psi}_{lm}({\bf r}_{j})] [{\bf e}_{k} \cdot {\bm \psi}_{lm}^{*}({\bf r}_{k})],
\nonumber\\
K_{jk}^{(l)}
&=\frac{4\pi}{2l+1} \sum_{m=-l}^{l} [{\bf e}_{j} \cdot {\bm \psi}_{lm}({\bf r}_{j})] [{\bf e}_{k} \cdot {\bm \chi}_{lm}^{*}({\bf r}_{k})],
\end{align}
where ${\bm \psi}_{lm}({\bf r})={\bm \nabla} \left [r^{-l-1}Y_{lm}(\hat{\bf r})\right ]$ and ${\bm \chi}_{lm}({\bf r})={\bm \nabla} \left [r^{l} Y_{lm}(\hat{\bf r})\right ]$, $Y_{lm}(\hat{\bf r})$ being spherical harmonics.  Naturally, only the dipole ($l=1$) term contributes to $\Gamma_{jk}^{r}$, while $\Gamma_{jk}^{nr}$ includes all angular momenta. From diagonal elements, single-dipole-NP rates can be easily recovered for normal ($s=\perp$) and parallel ($s=\parallel$) orientations with respect to the NP surface \cite{nitzan-jcp81}: $\Gamma_{s}^{r}=\Gamma_{0}^{r}\left |1+a_{s}\alpha_{1}/r_{0}^{3}\right |^{2}$ and $\Gamma_{s}^{nr}= \left (3\Gamma_{0}^{r}/2k^{3}\right ) \sum_l b_{s}^{(l)} \alpha''_l /r_{0}^{2l+4}$, where $a_{\perp}=2$, $b_{\perp}^{(l)}=(l+1)^2$, and $a_{\parallel}=-1$,  $b_{\parallel}^{(l)}=l(l+1)/2$. 

Radiated energy in the unit frequency interval is obtained by integrating spectral intensity over solid  angle, $dW/d\omega= (c\epsilon_{0}/4\pi^{2})\int \left |{\bf E}({\bf r},\omega)\right |^{2}r^{2}d\Omega$, and averaging the result over initial phases of oscillators, $\varphi_{j}$. Here the \emph{far-field} ${\bf E}({\bf r},\omega)$ is given by Eq.~(\ref{electric}), where $v_{j}$ is the solution of Eq.~(\ref{displ}) and ${\bf G}({\bf r},{\bf r}_{j},\omega)$ is  the \emph{large} $r$ asymptotics of the Mie Green diadic \cite{ruppin-jcp82}. The details will be given elsewhere \cite{to-be}, and the final result reads
\begin{equation}
\label{spectral}
\frac{dW}{d\omega}= \frac{1}{4\pi}{\rm Tr} \biggl[\frac{ \sqrt{\epsilon_0}\hbar \omega_0}{\bigl (\omega-\omega_{0}  -  \hat{\Sigma}\bigr )\bigl (\omega-\omega_{0}  -  \hat{\Sigma}^{\dagger}\bigr )} \,\hat{\Gamma}^{r} \biggr].
\end{equation}
In the \emph{absence} of dipole coupling, i.e., for purely diagonal $\Sigma_{jk}=-i\delta_{jk}\Gamma$ and $\Gamma_{jk}^{r}=\delta_{jk}\Gamma^{r}$, the frequency integration recovers radiated energy of $N$ \emph{isolated} dipoles near a NP, $W=N(\sqrt{\epsilon_0}\hbar \omega_0/4) Q=NW_{0}$. 

To illustrate the effect of SP coupling between emitters, first consider $N$ dipoles randomly distributed in a solid angle around a NP at the \emph{same} distance $d\gtrsim R$ from its surface, with normal or parallel orientations. At such distances, the high angular momenta ($l>1$) contributions to $\Gamma_{jk}^{nr}$ are suppressed, and decay matrices Eq.~(\ref{width}) take simple form $\Gamma_{jk}^{r}=\Gamma_{s}^{r}  A_{jk}$ and $\Gamma_{jk}^{nr}=\Gamma_{s}^{nr}  A_{jk}$, where $A_{jk}={\bf e}_{j}\cdot {\bf e}_{k}$ is cosine matrix, and $\Gamma_{s}^{nr} $ includes only $l=1$ term ($s=\perp,\parallel$). We now introduce \emph{cooperative decay} matrices as $\gamma_{\mu\nu}^{r}=(N\Gamma_{s}^{r}/3)  B_{\mu\nu}$ and $\gamma_{\mu\nu}^{nr}=(N\Gamma_{s}^{nr}/3)  B_{\mu\nu}$, where $B_{\mu\nu}=(3/N)\sum_{j} e_{j\mu}e_{j\nu}$ is $3\times 3$  matrix in coordinate space with ${\rm Tr}\hat{B}=3$. Now we note that, since ${\rm Tr}\hat{A}^{n}={\rm Tr}(N\hat{B}/3)^{n}$ for any integer $n$, the $N\times N$ matrices $\Gamma_{jk}^{r,nr}$ have only \emph{three non-zero eigenvalues} coinciding with those of matrices $\gamma_{\mu\nu}^{r,nr}$. Therefore, only these eigenvalues contribute to the spectral function,
\begin{align}
\label{spectral1}
\frac{dW}{d\omega}
&= \frac{\sqrt{\epsilon_0}\hbar \omega_0}{4\pi} \sum_{\mu=1}^{3} \frac{Q_{\mu}\gamma_{\mu}}{(\omega-\omega_{0})^2 + \gamma_{\mu}^2},
\\
\label{efficiency}
Q_{\mu}
&=\frac{\gamma_{\mu}^{r}}{\gamma_{\mu}}=
\frac{\Gamma_{s}^{r}}{\Gamma_{s}^{nr}+\Gamma_{s}^{r}+ (3\Gamma_{0}^{nr}/N\lambda_{\mu})}
\end{align}
where $\gamma_{\mu}^{r}=\lambda_{\mu} N \Gamma_{s}^{r}/3$ and $\gamma_{\mu}= \lambda_{\mu} N \left (\Gamma_{s}^{r}+\Gamma_{s}^{nr}\right )/3 +\Gamma_{0}^{nr} $ are radiative and total decay rates of \emph{plasmonic SR states}, $Q_{\mu}$ are their quantum efficiencies, and $\lambda_{\mu}\sim 1$ are eigenvalues of $B_{\mu\nu}$. Importantly, \emph{both} radiative and nonradiative rates of SR states are enhanced by factors $\sim N/3$ (for each degree of freedom). However, these factors effectively \emph{cancel out} in the quantum efficiencies, $Q_{\mu}$.  Furthermore, the contribution of $\Gamma_{0}^{nr}$ is the denominator of Eq.~(\ref{efficiency}) is suppressed by the factor $N^{-1}\ll 1$, i.e., $Q_{\mu}$ are not sensitive to intramolecular relaxation processes. Not too far from a NP, when $\Gamma_{0}^{nr}\ll (\Gamma_{s}^{r}+\Gamma_{s}^{nr})$, SR and single-molecule efficiencies essentially \emph{coincide},  $Q_{\mu}\simeq Q$. Integrating Eq.~(\ref{spectral1}) over frequency, we obtain Eq.~(\ref{energy}).

The origin of three plasmonic SR states is that even though dipole orientations may be uniform with respect to the curved metal surface (e.g., normal to it), they are not uniform in space, and vice versa. Note that for purely radiative coupling and for uniform dipole orientations, there is only a single SR state \cite{dicke-pr54,andreev-book,scheibner-naturephys07}.
  \begin{figure}
  \centering
  \includegraphics[width=0.8\columnwidth]{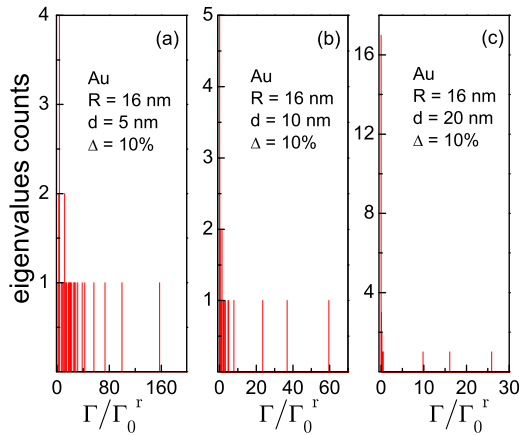}
  \caption{\label{fig:distr-random} (Color online) Distribution of decay rates for 30 dipoles around Au NP at several average (with  10\% fluctuations) distances to its surface.}
  \end{figure}

\emph{Numerical simulations and discussion}---Although Eqs.~(\ref{spectral1},\ref{efficiency}) were derived for moderate distances, $d\gtrsim R$,  these results apply even close to NP surface.  In Fig.~\ref{fig:distr-random}, we show numerical simulations of the eigenvalue distribution of $\Gamma^{r}_{jk}+\Gamma^{nr}_{jk}$ for 30 molecules with normal dipole orientations randomly placed in spherical angle around Au NP of radius $R=16$ nm (in a medium with $\epsilon_{0}=1.77$). Usually, stretching and folding of linker molecules causes fluctuations of dipole-to-surface distances by some amount $\Delta$ around the average value $d$ \cite{feldmann-prl02,lakowicz-jf02,feldmann-nl05}, so 10\% spread in distances ($\Delta/d =0.1$) was included. Calculations were performed at SP energy of 2.31 eV,  and NP polarizabilities $\alpha_{l}(\omega)$ with up to $l=30$ in Eq.~(\ref{width}) were included in $\Gamma^{nr}_{jk}$. To account for quantum-size effects in small NP, we incorporated in $\alpha_{l}(\omega)$ the Landau damping of $l$-pole plasmons, characterized by rate $\gamma_{l}\simeq 3lv_{F}/4R$, where $v_{F}$ is the Fermi velocity in the metal. The distribution of decay rates reveals the increasing role of nonradiative processes as the average distance to NP surface is reduced (see Fig.~\ref{fig:distr-random}). For $d=20$ nm, there are only \emph{three} non-zero eigenvalues corresponding to SR states, in agreement with Eq.~(\ref{spectral1}). With decreasing $d$, the remaining $N-3$ sub-radiant states start emerging  ($d=10$ nm) and, at small distances ($d=5$ nm), all system eigenstates acquire a finite decay rate. Note that a similar result holds for any dipole orientations; e.g., for random orientations in tangent plane, the three SR modes in Eq.~(\ref{spectral1}) are well separated from the rest.
\begin{figure}[tb]
  \centering
  \includegraphics[width=0.8\columnwidth]{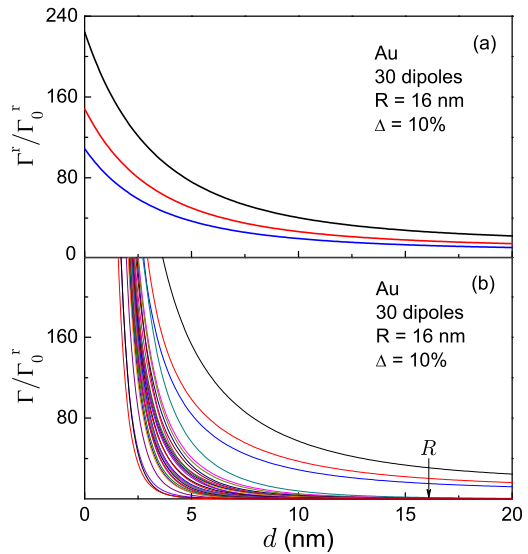}
  \caption{\label{fig:decay-distance} (Color online) Eigenvalues of (a) radiative, $\Gamma^{r}_{jk}$, and (b) full, $\Gamma_{jk}$,  decay matrices vs. average distance to NP surface for 30 dipoles randomly distributed around Au NP.}
  \end{figure}
  \begin{figure}[tb]
  \centering
  \includegraphics[width=0.8\columnwidth]{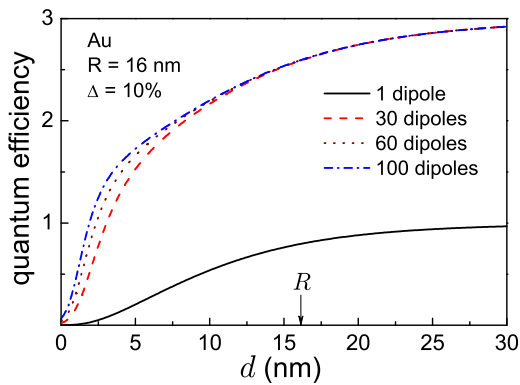}
  \caption{\label{fig:efficiency} (Color online) Combined q   uantum efficiencies for ensembles of 30, 60, and 100 dipoles compared to that for an isolated dipole near Au NP.}
  \end{figure}

The crucial distinction between the photonic and plasmonic Dicke effects stems from \emph{non-radiative coupling} between emitters in the latter. In the absence of a NP, the system eigenstates are eigenvectors of \emph{radiative} decay matrix, $\hat{\Gamma}_{0}^{r}$, and represent super- or sub-radiant modes characterized by strength of their coupling to the radiation.  In the presence of a NP,  optically bright and dark states are also defined through $\hat{\Gamma}^{r}$, Eq.~(\ref{width}), whose eigenvectors describe three SR and $N-3$ sub-radiant modes \emph{regardless} of dipole-surface separation, as shown in Fig.~\ref{fig:decay-distance}(a). However, the \emph{true} system eigenstates, $| j\rangle$, are described by the \emph{full} decay matrix, $\hat{\Gamma}^{r}+\hat{\Gamma}^{nr}$, with eigenvalues $\Gamma_{j}$, whose nondiagonal elements include non-radiative coupling. Therefore, the true radiative decay rates are given by expectation values $\Gamma_{j}^{r}=\langle j|\hat{\Gamma}^{r}| j\rangle $, while the quantum efficiencies are $Q_{j}=\Gamma_{j}^{r}/\Gamma_{j}$. In the case, e.g., of normal or tangential orientations, the dipole terms in the non-radiative decay matrix Eq.~(\ref{width}) possess the same symmetry as the radiative decay matrix, so for distances not too close to the NP, the SR modes are still system eigenstates. At the same time, the high-$l$ terms in $\hat{\Gamma}^{nr}$ have different symmetry than $\hat{\Gamma}^{r}$ and, therefore, they cause \emph{mixing} of SR and sub-radiant modes [see Fig.~\ref{fig:decay-distance}(b)]. However, except for very small distances, this mixing is \emph{weak} so that the emission remains cooperative. In Fig.~\ref{fig:efficiency}, we compare distance dependence of \emph{combined} quantum efficiencies, $Q_{ens}=\sum_{j}Q_{j}$, for ensembles of 30, 60, and 100 molecules to the single-molecule $Q$ near a $R=16$ nm gold NP. For distances $d\gtrsim R/2$ (8 nm), all \emph{ensemble} dependences collapse into single curve with amplitude $3Q$ (we used $\Gamma_{0}^{nr}=1.08\times 10^{9}$ s$^{-1}$ for Cy5), indicating that the emission is dominated by SR modes. Even closer to NP surface, up to $d\approx 5$ nm, the emission remains cooperative, although deviations from $3Q$ behavior appear. For smaller $d$, the eigenstates are no longer SR and sub-radiant modes, and cooperative emission is destroyed by non-radiative processes. 

The above analysis holds when the overall system size is smaller than the radiation wavelength.  This condition for cooperative emission also allowed us to use the long-wave approximation for the Mie theory Green diadic, where we disregarded the real part containing direct dipole-dipole interactions between the emitters. In absence of a NP, the latter can be considerably stronger than radiative coupling and may lead to a suppression of the photonic Dicke effect for large ensembles \cite{andreev-book}. In contrast, in the presence of a NP and close to the SP resonance,  the self-energy matrix is dominated by its imaginary part, while the dipole-dipole interactions lead to a spread of dipole frequencies around average value $\omega_{0}$ \cite{mukamel-jcp89,stockman-prl97}. However, a such a disorder affects only sub-radiant modes by removing the degeneracy in their spectral positions, but has no significant effect on SR modes \cite{shahbazyan-prb00}. Therefore, the plasmonic Dicke effect can survive the dipole-dipole interactions even for large ensembles.

Finally, the predicted plasmonic Dicke effect could be observed in experiments with controllable separation of emitters from the NP surface. In the recent experiment on Cy5 dyes linked with an Au NP \cite{feldmann-nl05}, a systematic study of distance dependence for the ensemble fluorescence was performed. Even though the number of emitters was not fixed, a fast saturation of quantum efficiency with increasing distance was observed, consistent with Eq.~(\ref{energy}).

This work was supported by NSF Grant No. DMR-0606509, by NIH  Grant No. 2 S06 GM008047-33, and by DoD contract No. W912HZ-06-C-0057.

{}

\end{document}